# Chiral excitations of magnetic droplet solitons driven by their own inertia

Morteza Mohseni,[1,†] D .R. Rodrigues,[2] M. Saghafi,[1] S. Chung,[3,4] M. Ahlberg,[4] H. F. Yazdi,[1] Q. Wang,[5] S.A.H. Banouazizi,[3] P. Pirro,[5] J. Åkerman,[3,4] and Majid Mohseni[1,*]

[1]Department of Physics, Shahid Beheshti University, Evin, Tehran 19839, Iran

[2]Institut für Physik, Johannes Gutenberg-Universität Mainz, D-55128 Mainz, Germany

[3]Materials and Nano Physics, School of Engineering Sciences, KTH Royal Institute of Technology, Electrum 229, 164 40 Kista, Sweden

[4]Department of Physics, University of Gothenburg, Fysikgränd 3, 412 96 Gothenburg, Sweden

[5]Fachbereich Physik and Landesforschungszentrum OPTIMAS, Technische Universität Kaiserslautern, 67663 Kaiserslautern, Germany

*Correspondence to: m-mohseni@sbu.ac.ir, majidmohseni@gmail.com.

†Current address: Fachbereich Physik and Landesforschungszentrum OPTIMAS, Technische Universität Kaiserslautern, 67663 Kaiserslautern, Germany

**Abstract:** The inertial effects of magnetic solitons play a crucial role in their dynamics and stability. Yet governing their inertial effects is a challenge for their use in real devices. Here, we show how to control the inertial effects of magnetic droplet solitons. Magnetic droplets are strongly nonlinear and localized autosolitons than can form in current-driven nanocontacts. Droplets can be considered as dynamical particles with an effective mass. We show that the dynamical droplet bears a second excitation under its own inertia. These excitations comprise a chiral profile, and appear when the droplet resists the force induced by the Oersted field of the current injected into the nanocontact. We reveal the role of the spin torque on the excitation of these chiral modes and we show how to control these modes using the current and the field.

Inertia measures how a physical object resists the accelerating forces applied to it. Within the inertial frame of reference, an object with a mass can undergo a restoring force, as in the case of a spring forced out of its equilibrium position. Indeed, inertial effects can be seen in almost all physical systems, from stars [1] through molecules [2], to solitons [3]. Solitons are self-localized wave packets that can form in a wide range of environments involving dispersion and nonlinearity, such as liquids, optics, plasma, Bose–Einstein condensates, and magnets [4–8]. The use of magnetic solitons such as skyrmions, domain walls, vortices, and droplets [8–14] opens many opportunities in data storage and communications technologies. However, stabilizing the dynamical solitons against forces and fluctuations remains a great challenge. For example, magnetic solitons exhibit inertial effects associated with their ability to store energy, for example by shape deformation [14-15]. In this picture, magnetic solitons can be treated as particles with an effective mass [3,15–18], although direct control of their inertial effects and realizing massless soliton motions is a key element of success in this field [18,19].

Magnetic droplets are localized autosolitons that can be formed using the spin transfer torque (STT) effect in layers with large perpendicular magnetic anisotropies (PMA) [20–22]. The injection of an electrical current into the nanocontact (NC) provides enough spin angular momentum (gain) to compensate for the viscous damping (dissipation) of the host magnet, while the PMA acts as a nonlinearity, canceling out the dispersion effect. This causes the propagating spin-waves to become modulationally instable, forming the droplet underneath the NC.

Due to their dynamical features and internal degrees of freedom, droplets can be considered dynamical particles carrying an effective mass [16,23], which they gain from the applied STT. This means that, in the absence of the STT, they lose their effective mass and dissipate [24,25]. Thus, direct observation and control of their inertial effects would constitute a major breakthrough.

Here, we report on the observation and control of the inertial effects of magnetic droplet solitons. Inertia is evidenced when the droplet resists the force induced by the Oersted field of the applied current. This force pushes the droplet outside the NC perimeter, while the droplet inertia restored by the STT opposes this force. This leads to the appearance of an excitation of two chiral modes in the droplet's precessional boundary. These chiral modes consist of clockwise and counterclockwise waves that form two eigenfrequencies in the magnetodynamic spectrum of the system. We illustrate how these modes can be controlled by the current and the field.

The system is an orthogonal spin torque nano-oscillator (STNO) that has a Co thin film (8 nm) as the fixed layer and a CoNi multilayer (3.6 nm) as the free layer. These are separated by an 8-nm Cu spacer (Fig. 1a) [26].

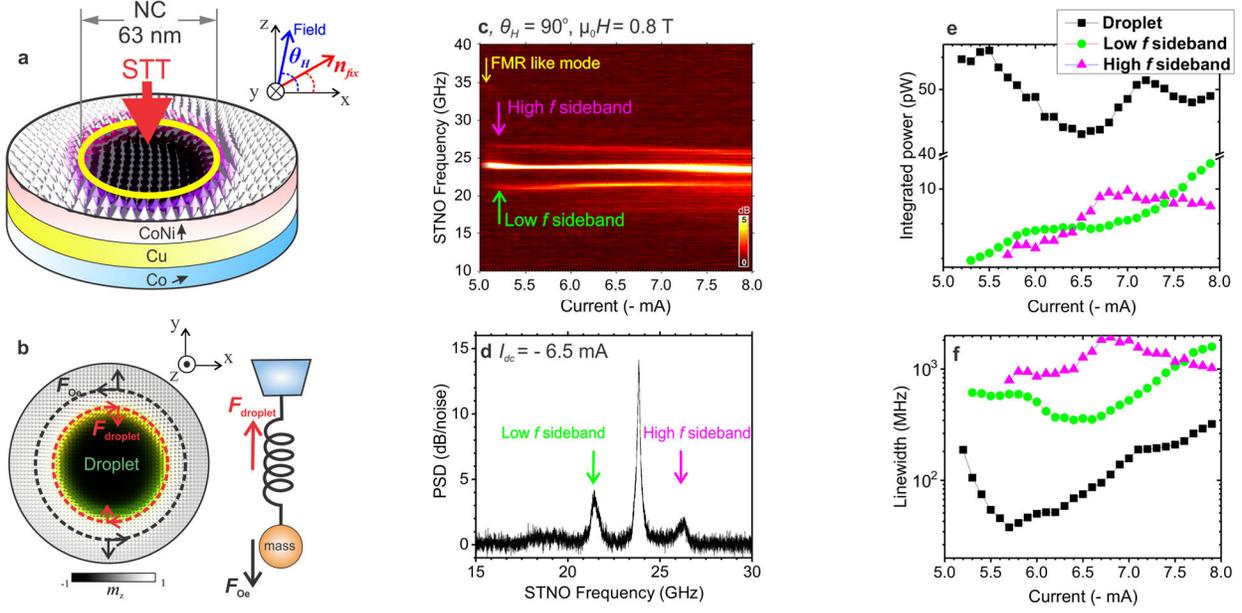

Figure 1. Schematic of the device structure and the current-dependent measurements: (a) structure of the orthogonal STNOs and droplet nucleation beneath the NC; (b) schematic of the force of the Oersted field $F_{Oe}$ against the restoring force of the droplet $F_{droplet}$ due to the STT, similar to a spring pulled out of equilibrium; (c) STNO frequency vs. applied current at $\theta_H = 90°$ and $\mu_0 H_{ex} = 0.8$ T; (d) STNO frequency spectrum extracted from (c) at $I_{dc} = -6.5$ mA; (e) and (f) show the integrated power and linewidth of the droplet mode and sidebands as a function of current, which corresponds to (c).

In the absence of an applied field, the magnetization of the fixed layer lies in the plane of the films, while the free layer is intrinsically out-of-plane due to its PMA. Under this configuration, we can alter the angle of the fixed layer magnetization using an external field applied perpendicularly to the film plane. A dc current is injected into the NC (63 nm in diameter) in order to provide enough spin angular momentum to drive the magnetization of the free layer and form the droplet.

The applied current generates an Oersted field inside and around the NC. This Oersted field produces an effective field gradient around the NC and can thus induce a net force on the droplet's dynamical precession ($F_{Oe}$ in Fig. 1b). If this force is sufficiently strong, the droplet accelerates and is expelled away from the NC and, in the presence of damping, loses its effective mass and dissipates [16,20]. However, in an opposite manner, the presence of the STT increases the effective mass of the droplet and generates an effective restoring force that centers the droplet under the NC ($F_{droplet}$ in Fig. 1b). These two competing forces in the plane of the NC recall a simple spring, which oscillates when pulled out of its equilibrium position.

We initially set the field to $\mu_0 H = 0.8$ T ($\theta_H = 90°$, perpendicular to the plane). The current-sweep frequency spectrum of the STNO is shown in Figure 1c. At an applied current of $I_{droplet} = -5.1$ mA, the frequency suddenly drops from the FMR-like mode and the integrated power increases dramatically: a clear sign that a droplet is nucleating beneath the NC [21,22]. The nucleated droplet has a maximum frequency of $f_{max} = 24.05$ GHz. The frequency decreases with increasing current—a red-shift trend that illustrates that a negative nonlinearity is required for soliton formation [20]. More importantly, two sidebands appear simultaneously with the droplet response, indicating automodulation effects in the system. In fact, this is the main evidence of the inertia of the droplet corresponding to the competing forces $F_{Oe}$ and $F_{droplet}$. Higher current densities provide a higher Oersted field inside and around the NC, and consequently the effective field becomes more inhomogeneous in the region where the droplet nucleates. This increases the force of the Oersted field ($F_{Oe}$). In addition, higher currents lead to an increase in thermal fluctuations in the system. Both mechanisms are expected to affect droplet instabilities [20,22,27,28]. However, our measurements show that the sideband frequencies slowly approach the main frequency with increasing current. This implies that the droplet becomes more stable on account of the higher STT provided by the current ($F_{droplet}$).

The frequency spectrum of the STNO at a current of $I_{dc}$ = - 6.5 mA is presented in Figure 1d. The figure shows three distinct peaks corresponding to the droplet and its sidebands. The frequency difference between the sidebands and the main peak is $\Delta f$ = 2.45 GHz. Figure 1e–f illustrates the evolution of the integrated peak power and linewidth with current. The power of the droplet mode decreases when the sidebands appear, since the energy provided is shared between the three different modes. Moreover, these dynamics reduce the droplet power as a portion of the droplet is periodically expelled from the NC region. The unstable dynamical motion also leads to a larger linewidth [22].

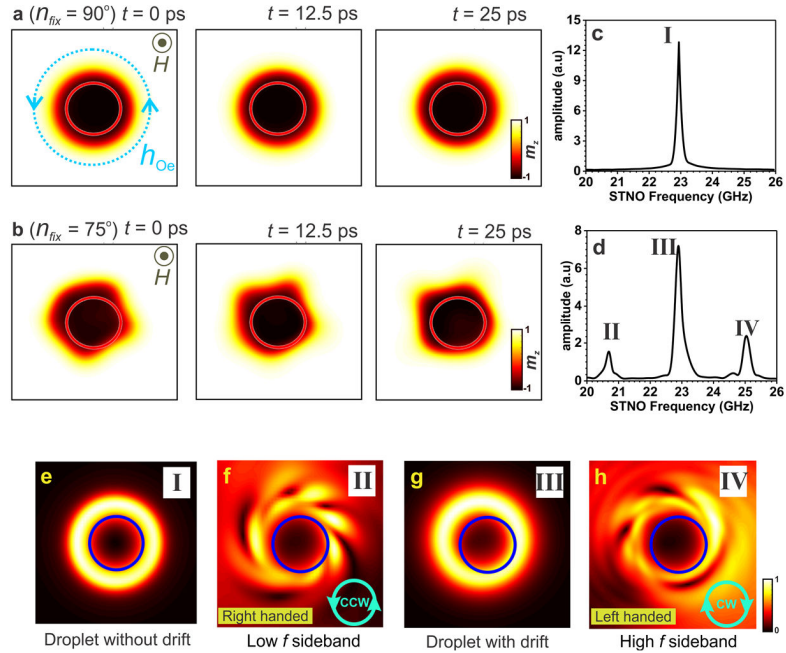

Figure 2. Results of micromagnetic simulations with the external field $\mu_0 H$ = 0.75 T applied out-of-plane ($\theta_H$ = 90°) and $I_{dc}$ = - 6.5 mA. Snapshots of the droplet dynamics when the fixed layer magnetization angle is (a) $n_{fix}$ = 90° and (b) $n_{fix}$ = 75°. The arrows in the green circle show the direction of the Oersted field; (c–d): frequency spectra given by the dynamics for $n_{fix}$ = 90° (shown in a) and for $n_{fix}$ = 75° (shown in b), respectively; (e–h): spatial profile of the modes corresponding to the peaks labeled by I, II, III, and IV in (c) and (d).

Micromagnetic simulations [26,29] reveal the existence of the excitation of chiral modes of the droplet and shed light on the mechanisms behind this drift dynamic. Figure 2a demonstrates that the droplet motion is completely homogenous when the fixed layer angle is out of plane, $n_{fix}$ = 90° (see also supplementary animation Movie S1). The dynamics correspond to a single peak in the frequency spectrum (Fig. 2c) and an entirely uniform spatial profile of the droplet mode (Fig. 2e).

We then tilt the fixed layer magnetization angle to, for example, $n_{fix} = 75°$, which is more similar to our experiments given the applied field is not large enough to saturate the fixed layer and rotates its magnetization angle to normal angles. This in fact, breaks the symmetry of the driving STT by providing an in-plane component for the spin torque. The results, shown in Figure 2b, indicate that the droplet is displaced to the left side of the NC and its perimeter continuously deforms (see supplementary animation Movie S2). This means that the droplet boundary is no longer fully uniform, and there must be a net force acting as a perturbation to the system. The frequency spectrum under this condition shows that the droplet dynamics exhibit three peaks, as shown in Figure 2d. Indeed, the main peak in Figure 2d is given by the principal droplet mode, and the two sidebands are related to the periodic deformation. The sideband frequency shift of $\Delta f$ ~ 2.2 GHz is similar to what was found in the experiments (Fig. 1d). Moreover, due to the drift of the droplet from the NC, the main peak is not fully symmetric.

The chirality of these excitations can be demonstrated by mapping their spatial profiles. Figure 2f–h reveals that the low $f$ modes and the high $f$ modes are given by opposite directions of precession: counterclockwise (CCW) and clockwise (CW) which follow the right-hand and left-hand rules, respectively. However, due to the presence of a noncollinear STT (tilted $n_{fix}$), the rotational symmetry of the droplet profile is broken. Further simulations demonstrate that the chiral modes depend on the direction of the Oersted field ($h_{Oe}$). If the Oersted field is inverted, the sense of rotation of the precession is also inverted [26]. In addition, the sidebands disappear and the droplet dislocation vanishes if $n_{fix} = 90°$, or if the Oersted field is set to zero. This means that the force induced on the droplet by the Oersted field ($F_{Oe}$) tends to displace the droplet outside of the NC. However, due to its internal inertia restored by the STT, the droplet resists this force ($F_{droplet}$), resulting in spatial deformation of the droplet's boundary. As a consequence, there is excitation of a chiral precession corresponding to two frequencies at the frequency spectrum of the system.

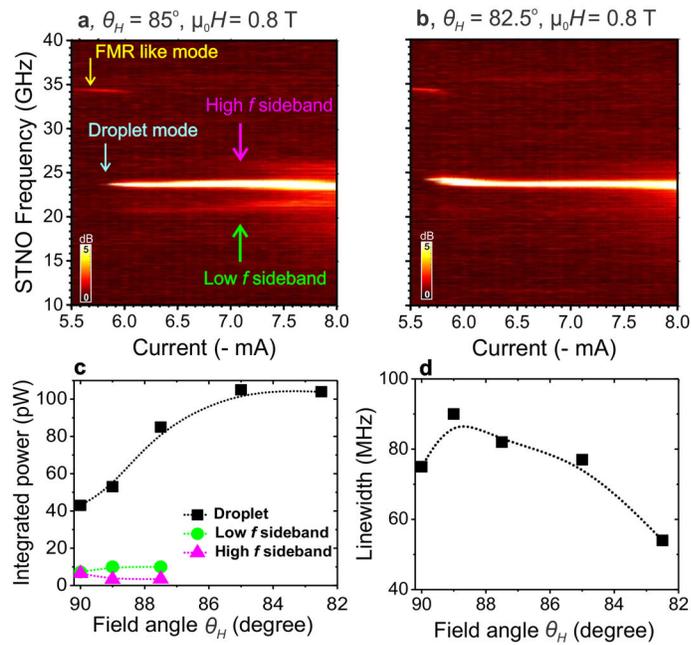

Figure 3. STNO frequency vs. applied current at canted fields when $\mu_0 H = 0.8$ T; (a): $\theta_H = 85°$; (b): $\theta_H = 82.5°$; (c) and (d): integrated power and linewidth of the droplet and the sidebands versus the field angle $\theta_H$, respectively. For (c) and (d), the $\mu_0 H_{ex} = 0.8$ T and $I_{dc} = -6.5$ mA; the lines are visual guides.

We now tilt the field angle to $\theta_H = 85°$ and $\theta_H = 82.5°$ and repeat the measurements as shown in Figure 3a–b. The nucleation current increases slightly at off-normal angles, which can be explained by the droplet's nucleation boundaries [30]. Obviously, the sideband intensities become weaker for smaller angles, while only the main peak is observed at $\theta_H = 82.5°$. This implies that the inertial effect associated with the sidebands is reduced, and eventually vanishes, when the field is tilted towards the film plane. This conclusion is also supported by Figure 3c–d, which shows the angular dependence of the integrated power and linewidth of the three modes at a current of $I_{dc} = -6.5$ mA and an applied field of $\mu_0 H = 0.8$ T, respectively. The droplet power rises substantially when the sidebands become vanishingly small around $\theta_H = 85°$. The available energy is shared by the modes at large angles, while all the energy is dissipated through the droplet precession at smaller angles, resulting in increased power. The disappearance of the chiral excitations ($\theta_H < 87.5°$) is also accompanied by a reduction in the main mode linewidth, which means that the droplet becomes more stable.

In fact, the presence of the canted fields breaks the spatial symmetry of the effective field around the NC, and hence the spatial symmetry of the droplet profile [31–33]. In addition, it reduces the angle of the fixed layer magnetization. We therefore conclude that a symmetric but an inhomogeneous effective field is required for sideband formation, in accordance with our earlier discussion on the all-perpendicular conditions (Figs. 1c and 2a).

Further measurements demonstrate that sidebands becomes weaker and finally disappear in the presence of higher fields [26]. This is caused by the presence of higher angle of the fixed layer magnetization $n_{fix}$, in the presence of higher fields. Moreover, the appearance of the chiral modes destroys the breathing modes of the droplet [34], implying that a new dynamic dominates the spatial deformation of the droplet profile. This is expected due to the excitation of the chiral precession [26].

We now quantitatively analyze the mechanism behind the chiral motions of the droplet based on the following. We describe the magnetization of the free layer by the following Free energy equation,

$$F = \tau \int d^2x (A(\nabla \boldsymbol{m})^2 + K(1 - m_z^2) + \mu_0 H_z M_s \hat{\boldsymbol{z}} \cdot \boldsymbol{m} + U_{dd}) \tag{1}$$

where $\boldsymbol{m} = \boldsymbol{M}/M_s$ is the unit magnetization field, $\tau$ is the thickness of the free layer, the constants $A$, $K$ are the exchange and anisotropy strengths, $H_z$ is the external field along the $\hat{\boldsymbol{z}}$ direction, and $U_{dd}$ corresponds to the stray field contribution. We consider that the major contribution from the stray fields is a rescaling of the anisotropy strength, $K_0 = K - \frac{\mu_0}{2} M_s^2$ where $\mu_0$ is the vacuum permeability, and any other contribution is a small perturbation and does not contribute significantly to the magnetization dynamics.

The dynamics of the free layer is then given by the Landau-Lifshitz-Gilbert-Slonczewski equation of motion [35–39],

$$\dot{\boldsymbol{m}} = -\gamma \boldsymbol{m} \times \boldsymbol{H}_{eff} + \alpha \boldsymbol{m} \times \dot{\boldsymbol{m}} + \gamma \beta \frac{\epsilon + \alpha \epsilon'}{1 + \alpha^2} \Theta(r_{NC} - r) \boldsymbol{m} \times (\boldsymbol{n}_{fix} \times \boldsymbol{m}) - \gamma \beta \frac{\epsilon' - \alpha \epsilon}{1 + \alpha^2} \Theta(r_{NC} - r) \boldsymbol{m} \times \boldsymbol{n}_{fix} + \gamma \frac{h_{Oe}}{r} \boldsymbol{m} \times \widehat{\boldsymbol{\psi}}. \tag{2}$$

where $\gamma = 1,76 \cdot 10^{11} rad/(sT)$ is the gyromagnetic ratio, $\beta = J_z\hbar/M_s e\tau$ is the spin torque coefficient where $\hbar$ is the reduced Planck constant, $J_z$ is the electric current density, $e$ is the elementary charge ($e > 0$), and $\epsilon = P\Lambda^2/((\Lambda^2 + 1) + (\Lambda^2 - 1)(\boldsymbol{m} \cdot \boldsymbol{n}_{fix}))$ where $P$ is the polarization, $\Lambda$ is the Slonczewski parameter, $\epsilon'$ is the secondary spin-torque parameter, $\Theta(r_{NC} - r)$ is the Heaviside step function with $r_{NC}$ the radius of the NC, $h_{Oe} = \mu_0 J_z/2\pi$ is the Oersted-field and $\hat{\boldsymbol{\psi}}$ is the angular tangential unitary vector. $\boldsymbol{n}_{fix}$ is the direction of the magnetization in the fixed layer which is set to $n_{fix} = (n_x, 0, n_z)$. By symmetry analysis, the Oersted field does not influence explicitly the dynamics of the precession or the average radius of the droplet. It is responsible, however, for a gyration motion of the soliton, corresponding to a rigid rotational motion. In contrast, the STT contribution in the area of the NC couples with any translation of the soliton.

The dynamics of droplets has been largely studied by considering linearized perturbations on a radially symmetric ansatz [23, 26, 40, 41]. In order to take into account the decay of the soliton due to the damping, we consider an approach usually applied to understanding the dynamics of skyrmions [42-44]. We consider the soliton a rigid texture and excitations modes are localized at the border of the soliton (Fig. 4), as observed in the experiment (Figs. 2e-h).

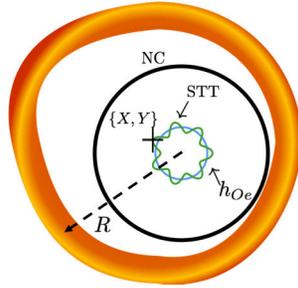

Figure 4: Sketch of the movement of the centre of the dynamical soliton based on Eq. 3. The orange ring corresponds to the border of the soliton. The black line represents the NC. The blue circle represents the gyration motion produced by the Oersted field. The oscillations corresponding to the green line come from the inertial effects created by the STT terms.

In this approach, the gyration motion is given by a motion of the center of the droplet $(X, Y)$ and other excitation modes are given in terms of $(\delta r, \delta \phi)$. Considering the rigid translational motion of the droplet given as $\delta r = X\cos\psi + Y\sin\psi$ and $\delta\phi = \frac{X}{r}\sin\psi - \frac{Y}{r}\cos\psi$ we obtain the following effective equations for the coordinates $X$ and $Y$,

$$\alpha D\dot{Y} = -\beta_1 f(r)(-Y\cos\phi - X\sin\phi)n_x + \beta_2\big((X\sin\phi f_1(R) - Y\cos\phi f_2(R))n_x + Yn_z f_3(R)\big)$$
$$-\gamma h_{Oe}\cos\phi\, sech(R)$$

$$\alpha D\dot{X} = -\beta_1 f(r)(X\cos\phi - Y\sin\phi)n_x + \beta_2\big((-Y\sin\phi f_1(R) - X\cos\phi f_2(R))n_x + Xn_z f_3(R)\big)$$
$$+\gamma h_{Oe}\sin\phi\, sech(R) \qquad (3)$$

where $D = (1/\pi)\int d^2x \partial_x \boldsymbol{m} \cdot \partial_x \boldsymbol{m}$ is the viscosity tensor, $\beta_1 = \gamma\beta\frac{\epsilon+\alpha\epsilon'}{1+\alpha^2}$ and $\beta_2 = \gamma\beta\frac{\epsilon'-\alpha\epsilon}{1+\alpha^2}$ are the strengths of the STT coupling, and $f(R), f_1(R), f_2(R)$, and $f_3(R)$ are functions depending on the radius of the soliton [26]. The lack of a topological charge, $Q = (1/4\pi)\int d^2x \boldsymbol{m} \cdot (\partial_x \boldsymbol{m} \times \partial_y \boldsymbol{m}) = 0$, one does not expect any chiral motion. However, due to the presence of inplane perturbations, one observes the coupling between the $X$ and $Y$ dynamical parameters. The unidirectional gyration motion produced by the Oersted field is damped by extra torques due to inplane components. They can be interpreted as inertial terms which means that the equations become of the sort $m\ddot{X} + \gamma\dot{X} + X = f(t)$. The equations above are rather hard to be solved and to obtain the exact motion of the center of the droplet. Due to the lack of a gyration term, one expects the formation of standing waves, in which the excitation mode frequencies are distributed symmetrically around the gyration mode, as observed in the numerical calculations, see Fig. 2d.

In conclusion, we have presented the observation and direct control of the inertial effects of magnetic droplet solitons. Inertia is evidenced when the droplet resist the force induced by the Oersted field injected into the nanocontact. This leads to an excitation of two chiral modes in the droplet's precessional boundary. We showed how to control these chiral modes using the current and the field. Our results (including ref. [22, 30, 31] and further measurements which are not shown) imply that this chiral excitation exists for NC diameters smaller than 100 nm and moderate out-of-plane fields (intermediate fixed layer magnetization angles), corroborating narrow conditions to observe and control their inertial effects. This is due to the fact that smaller droplets carry smaller effective masses and undergo stronger drifts in the presence of relatively similar forces. Controlling these chiral excitations may open up new approaches to precisely controlling the inertial effects of magnetic solitons. Furthermore, it complements the knowledge of the mechanisms that lead to inertial instabilities of magnetic solitons.

Financial support by the Deutsche Forschungsgemeinschaft (SFB/TRR 173 "Spin+X", Project B01) and by the DFG Priority Programme "SPP2137 Skyrmionics" is gratefully acknowledged. J. Åkerman acknowledges funding from The Swedish Research Council and the Knut and Alice Wallenberg Foundation. D.R. acknowledge funding from the German Research Foundation (DFG), projects EV 196/2-1, EV196/5-1 and SI1720/4-1, SFB/TRR 173 "Spin+X", Project A03 and as well the Emergent AI Center funded by the Carl-Zeiss-Stiftung. The authors acknowledges contributions by S. R. Sani for part of the device fabrication.